\documentclass[aps,preprint,showpacs,nofootinbib]{revtex4}
\usepackage{amssymb}
\usepackage{graphicx}
\usepackage{epsfig}

\begin{document}

%=====================================
%       put your definitions here
%=====================================
\def\beq{\begin{eqnarray}}
\def\eeq{\end{eqnarray}}
\def\non{\nonumber}
\def\la{\langle}
\def\ra{\rangle}
\def\Un{{\cal U}}

\def\pr{{Phys. Rev.}~}
\def\prl{{ Phys. Rev. Lett.}~}
\def\pl{{ Phys. Lett.}~}
\def\npb{{ Nucl. Phys. B}~}
\def\epjc{{ Eur. Phys. J. C}~}

%======================================================================

\title{ Interpretation of the ``$f_{D_s}$ puzzle" in SM and beyond }

\author{
  Zheng-Tao Wei$^{1}$\footnote{Email: weizt@nankai.edu.cn},
  Hong-Wei Ke$^{2}$\footnote{Corresponding author, email:khw020056@hotmail.com}, and
  Xiao-Feng Yang$^{1}$ }

\affiliation{
  $^{1}$Department of Physics, Nankai University, Tianjin 300071, China \\
  $^{2}$Department of Physics, Tianjin University, Tianjin 300072, China }

\begin{abstract}

\noindent The recent measurement on the decay constant of $D_s$
shows a discrepancy between theory and experiment. We study the
leptonic and semileptonic decays of $D$ and $D_s$ simultaneously
within the standard model by employing a lightfront quark model.
There is space by tuning phenomenological parameters which can
explain the ``$f_{D_s}$ puzzle" and do not contradict  other
experiments on the semileptonic decays. We also investigate the
leptonic decays of D and $D_{s}$ with a new physics scenario,
unparticle physics. The unparticle effects induce a constructive
interference with the standard model contribution. The nontrivial
phase in unparticle physics could produce direct CP violation which
may distinguish it from other new physics scenarios.

\end{abstract}


\maketitle

\section{Introduction}

Thanks to the recent experimental improvements from B factories,
BES, CLEO-c and hadron colliders, charm physics enters into a
``golden time" \cite{Bigi:2009df}. Many new charmed resonances are
observed, such as $D_{sJ}$ and X, Y, Z states. They open a new
window to study nonperturbative QCD. The last neutral meson mixing,
$D^0-\bar{D^0}$ mixing is observed at about $1\%$ level. This is a
typical flavor-changing-neutral-current (FCNC) transition which is
loop suppressed in the standard model (SM). It is expected that the
FCNC process is sensitive to new physics beyond the SM. On the
contrary, the leptonic decays of charm meson, e.g. $D_{(s)}\to
l\nu$, are tree dominated. New physics in these processes, if
exists, should be very small. However, recent measurements on the
leptonic decays of $D_s$ show that their ratios are larger than
expectation. In \cite{Rosner:2008yu}, the authors reviewed the
experimental and theoretical status of the decay constants of $D$
and $D_s$. For most approaches to calculate the decay constant of
$D_s$, theory predictions are smaller than the experiment. In
particular, there is a 3 standard deviation between the lattice QCD
calculation which claims a precise prediction \cite{Follana:2007uv}
and the experimental data. A recent updated QCD sum rules analysis
provide an upper bound which does not reduce the tension between
theory and experiment \cite{Khodjamirian:2008xt}. The above
discrepancy is sometimes called the ``$f_{D_s}$ puzzle".

To fully understand the $f_{D_s}$ puzzle, it requires an accurate
knowledge of strong interaction which is difficult to obtain at
present. In this study, we explore the problem from a
phenomenological point of view. Although the treatment has the
disadvantage that theoretical errors are not well under  control, it
permits an analytical treatment and its validity can be tested by
many processes. Our method is a light-front approach
\cite{Jaus:1999zv,Cheng:2003sm}. This is a relativistic quark model
and its essential element is hadron's light-front wave function. As
a successful phenomenological model, it has been widely applied to
calculate many different meson decay constants and form factors.
Within this approach, the $f_{D_s}$ puzzl problem really exists. In
a previous result \cite{Cheng:2003sm}, the decay constant of $D_s$
is $230$ MeV, which is quite different from the experimental value
of about $270$ MeV. How to explain the puzzle of decay constant
within the light-front approach  is our first task.

Another possible mechanism to explain the decay constant puzzle is
to introduce new physics effects beyond the SM. There have been many
scenarios proposed, e.g. charged Higgs and/or leptoquark
\cite{Akeroyd:2007eh, Dobrescu:2008er, Benbrik:2008ik,
Akeroyd:2009tn}. In order to explain the experiment, the new physics
effects must interfere constructively with the dominant SM
contribution. The charged Higgs model in \cite{Akeroyd:2007eh} is
excluded due to its destructive effect. Here, we suggest a new
physics scenario, unparticle physics
\cite{Georgi:2007ek,Georgi:2007si}. The unparticle is a
scale-invariant stuff with no fixed mass. The scale dimension of the
unparticle is in general fractional rather than an integral number.
Many unusual phenomena caused by unparticle are explored
\cite{Cheung:2007zza, Unparticle}. The non-trivial phase in
unparticle theory could induce constructive interference and even a
sizable CP violation. Some unparticle physics effects in the
leptonic decays of $B\to l\nu$ are studied in \cite{Huang:2007ax,
Zwicky:2007vv}. A detailed analysis of $D_{(s)}\to l\nu$ decays in
unparticle physics will be provided in this study.

If there exits new physics in subprocess $c\to sl\nu$ transitions,
it should also occur in semi-leptonic decays of charm mesons, such
as $D\to Kl\nu,~ D_s\to \eta,\eta'$ etc.. In most literature, the
semileptonic decays are either less considered or studied
separately. Simultaneously studying the charm leptonic and
semileptonic decays may help to discriminate different new physics
models. At the same time, to test the validity of light-front
approach, study of the semileptonic decays is necessary.

The paper is organized as follows: In Sec.II, the leptonic and
semileptonic decays of $D$ and $D_s$ mesons in SM are given. The
decay constants and the form factors are calculated within the
light-front approach. In Section III, we give an analysis of
leptonic decays in unparticle physics, concentring on the unparticle
effects on branching ratios and direct CP violation. The last
section is devoted to discussions and conclusions.

\section{Leptonic and semileptonic decays of $D$ and $D_s$ mesons in SM}

\subsection{$D_{(s)}^+\to l^+\nu$ decays in SM }

\begin{figure}[!htb]
\begin{center}
\begin{tabular}{cc}
\includegraphics[width=7cm]{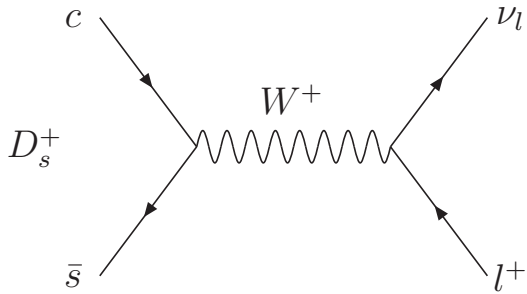}
\end{tabular}
\end{center}
\caption{ The lowest order diagram for the decay of $D_s^+\to
l^+\nu_l$ in SM.} \label{fig1}
\end{figure}

In SM, the purely leptonic decay of $D_{(s)}\to l\nu$  occurs via
annihilation of quark pair to a charged lepton and neutrino through
exchange of a virtual $W$ boson. The lowest order contribution is a
tree diagram which is depicted in Fig. \ref{fig1} for the decay of
$D_s^+\to l^+\nu_l$.  The effective Hamiltonian for subprocess $c\to
ql\nu$ transition at quark level is
 \beq
 {\cal H}_{\rm eff}^{SM}=\frac{G_F}{\sqrt 2}V_{cq}(\bar q c)_{V-A}
 (\bar\nu l)_{V-A},
 \eeq
where $q$ denotes $d$ for $D^+$ and $s$ for $D_s^+$ meson,
respectively, and $V-A=\gamma_\mu(1-\gamma_5)$. The weak radiative
corrections are so small that they can be safely neglected. The
strong interactions between $c$ quark and antiquark $\bar q$ by
exchange of gluons are incorporated in the definition of the decay
constant as
 \beq
 \la 0| \bar q \gamma_\mu\gamma_5 c |D(p)\ra=if_D p_\mu.
 \eeq
Here $D$ represents $D^+$ or $D_s^+$ to simplify the representation.

The decay width of $D\to l\nu$ is obtained straightforwardly by
 \beq
 \Gamma^{\rm SM}(D\to l\nu)=\frac{G_F^2|V_{cq}|^2}{8\pi}f_D^2
  M_D m_l^2\left(1-\frac{m_l^2}{M_D^2}\right)^2.
 \eeq
Because of helicity suppression for spin-0 particle decaying into
two spin-1/2 fermions, the decay rate is proportional to the lepton
mass square $m_l^2$.

It seems that the leptonic decays of $D_{(s)}\to l\nu$ are
theoretically simple and clean in SM. The weak interaction is well
determined and the information of strong interaction is encoded in
terms of decay constants of meson. If the CKM parameters are known,
the decay constants of $f_{D_{(s)}}$ can be extracted from the
measured decay ratios by using the above equation. From the recent
experimental data $Br(D^+\to \mu^+\nu_\mu)=(4.4\pm 0.7)\times
10^{-4}$ and $Br(D_s^+\to \mu^+\nu_\mu)=(6.2\pm 0.6)\times 10^{-3}$
\cite{PDG08}, we obtain
 \beq \label{eq:fexp}
 f_{D}^{\rm exp}=221\pm 17~ {\rm MeV}, \qquad \qquad
 f_{D_s}^{\rm exp}=270\pm 13~ {\rm MeV}.
 \eeq
where the CKM parameters $V_{cd}=-0.2257$ and $V_{cs}=0.9737$ are
used.

Recently, HPQCD and UKQCD collaborations improved their techniques
of lattice QCD (LQCD). They give a precise prediction for decay
constants of $D$ and $D_s$ by \cite{Follana:2007uv}
 \beq \label{eq:fLQCD}
 f_{D}^{\rm LQCD}=207\pm 4~ {\rm MeV}, \qquad \qquad
 f_{D_s}^{\rm LQCD}=241\pm 3~{\rm MeV}.
 \eeq
The most impressive thing about the above results is that the theory
error is very small, only 2\% or even smaller for $f_{D_s}$.
Comparing Eqs. (\ref{eq:fexp}) and (\ref{eq:fLQCD}), the experiment
is consistent with theory within the errors for $f_D$; while the
experiment data differ from theory prediction by about $3\sigma$
deviations for $f_{D_s}$. Since the $s$ quark in $D_s$ is heavier
than the $d$ quark in $D$ meson, the calculation for $f_{D_s}$ is
expected to be more reliable than $f_D$. The above discrepancy leads
to so called $f_{D_s}$ puzzle.

At first, we study whether the discrepancy between theory and
experiment can be reduced within the framework of SM. Our approach
is a covariant light-front quark model (LFQM)
\cite{Jaus:1999zv,Cheng:2003sm}. This is a relativistic quark model
in which a consistent and fully relativistic treatment of quark
spins and the center-of-mass motion can be carried out. This model
has many advantages. For example, the light-front wave function is
manifestly Lorentz invariant as it is expressed in terms of the
internal momentum fraction variables which is independent of the
total hadron momentum. Some applications of this approach can be
found in \cite{Hwang:2006cua}.

In the LFQM, the decay constant of a pseudoscalar meson is
represented by
 \beq \label{eq:fP}
 f_P=\frac{\sqrt{2N_c}}{8\pi^3}\int dx d^2 k_{\bot}
 \frac{A}{\sqrt{A^2+k_\bot^2}}\phi_P(x,k_{\bot}).
 \eeq
where $N_c=3$ is the color number of QCD, $A=m_1x+m_2(1-x)$ and
$m_{1,2}$ represent masses of constitute quark and antiquark in the
meson. The variable $x$ denotes the light-front momentum fraction
and $k_\bot$ denotes the intrinsic transverse momentum of the quark.
$\phi_P(x,k_{\bot})$ is the hadron light-front wave function. In
phenomenology, a Gaussian-type wave function is widely chosen as
 \beq
 \phi(x,k_{\bot})=N\sqrt{\frac{dk_z}{dx}}~{\rm exp}\left(
  -\frac{k_{\bot}^2+k_z^2}{2\beta^2} \right).
 \eeq
where the normalization constant $N=4(\frac{\pi}{\beta^2})^{3/4}$
and $\frac{dk_z}{dx}=\frac{e_1e_2}{x (1-x) M_0}$ with
$e_i=\sqrt{m_i^2+k_\bot^2+k_z^2}$ and
$M_0=\sqrt{\frac{k_\bot^2+m_1^2}{1-x}+\frac{k_\bot^2+m_2^2}{x}}$.
The parameter $\beta$ in the wave function determines the
confinement scale and is expected to be of order $\Lambda_{\rm
QCD}$. The quark masses and $\beta$ are the only required
phenomenological parameters which makes LFQM predictive.

\begin{figure}
\begin{center}
\begin{tabular}{cc}
\includegraphics[scale=0.6]{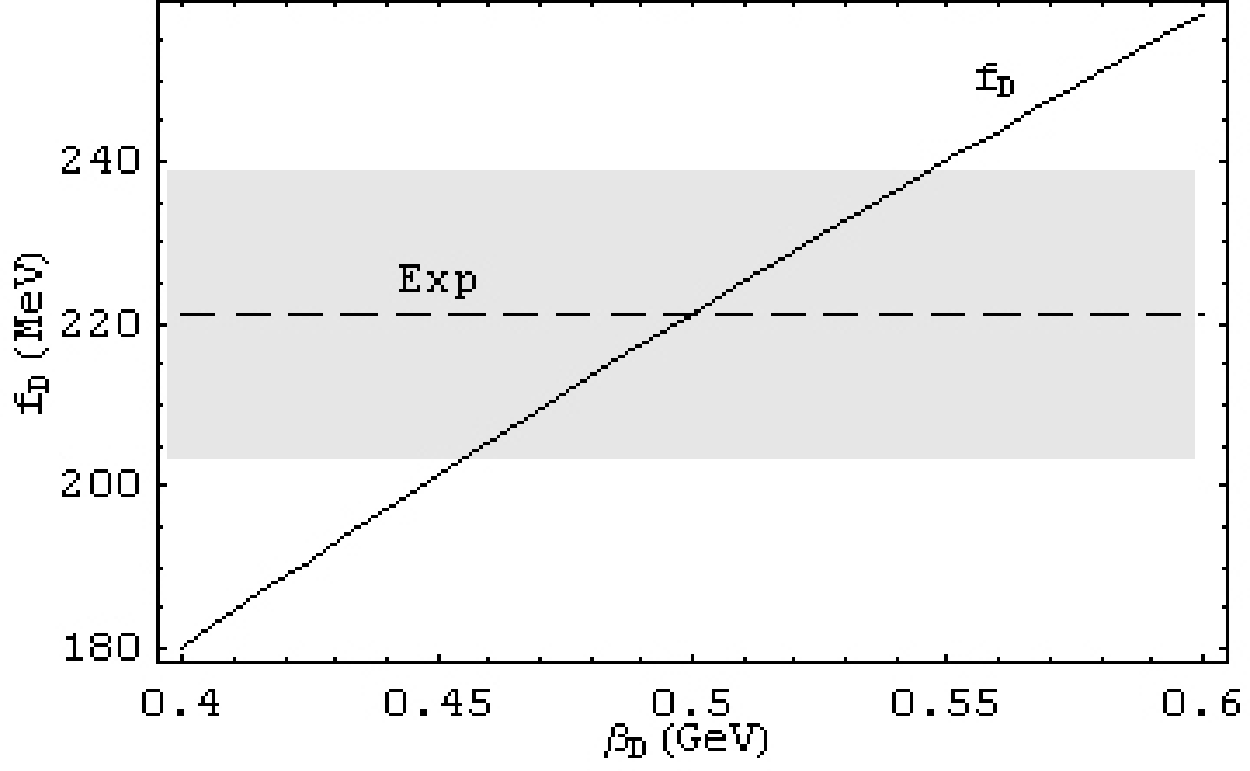}
\includegraphics[scale=0.6]{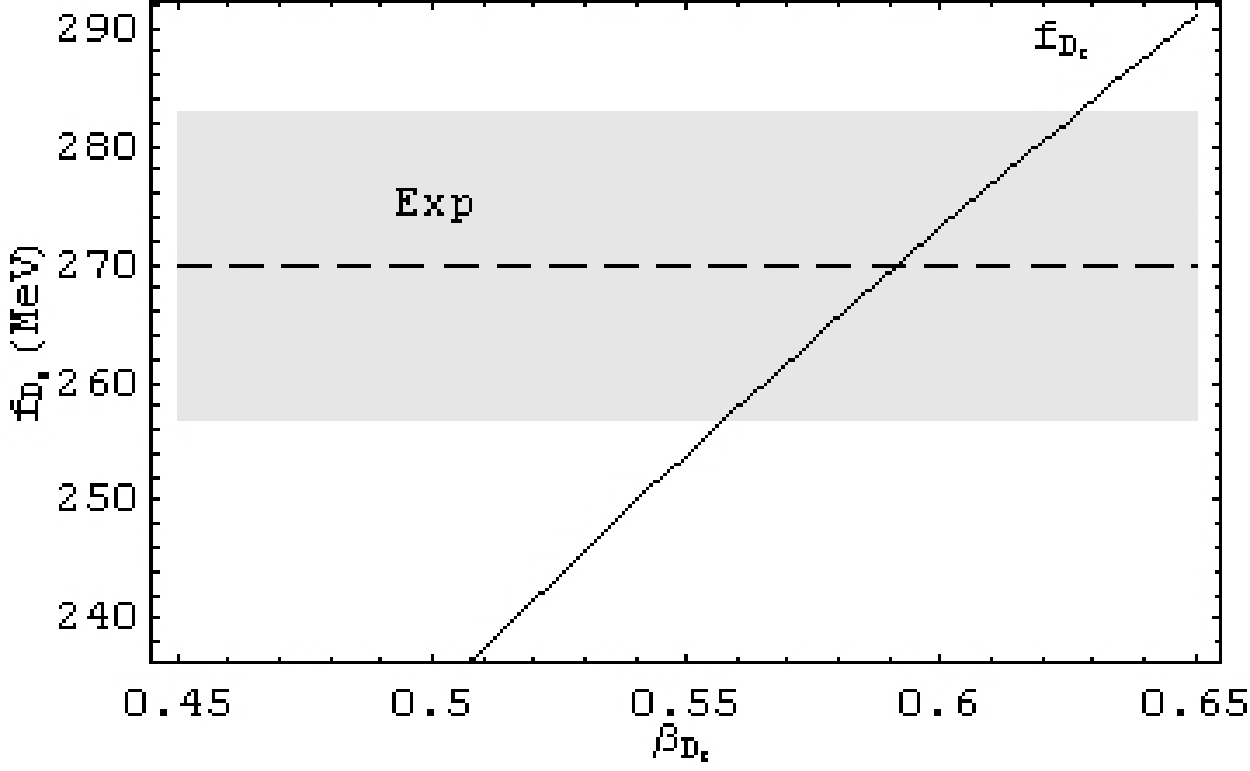}
\end{tabular}
\end{center}
\caption{ The decay constants $f_D$ and $f_{D_s}$ vs the parameter
$\beta$. The dashed line represents the central value and the gray
region the error of the experimental data.} \label{fig:fbeta}
\end{figure}

In this study, the constitute quark masses are chosen as in
\cite{Cheng:2003sm}: $m_d=0.26$ GeV and $m_c=1.40$ GeV. So the decay
constant depends only on parameter $\beta$. Figure\ref{fig:fbeta}
plots the decay constants of $f_D$ and $f_{D_s}$ that vary with
$\beta$. We can see that the decay constant increases as $\beta$
increases, and the relation is nearly a linear function. The slopes
$\partial f/\partial \beta$ for $D$ and $D_s$ are the same to a good
accuracy. It is noted that we observe this linear relation for the
first time although the reason for the relation is unknown. By
tuning the parameter $\beta$, it is easy to make theory be
consistent with experiment. The $\beta$ is within a reasonable
parameter region and its variation is small, less than 10\%. For
example, $\beta_{D_s}=0.59\pm 0.03$ GeV. In fact, as we will show
for the semileptonic decays, the recent accurate data provide better
predictions and determinations of parameters than before.

\begin{table}
\caption{\label{tab:fLFQM} The decay constants of $D$ and $D_s$ in
the light-front quark model (in units of MeV).}
\begin{ruledtabular}
\begin{tabular}{cccc}
                     & Model I  & Model II & Exp.               \\ \hline
 $f_D$ (MeV)         & 200      & 221      & $221\pm 17$        \\
 $f_{D_s}$ (MeV)     & 230      & 270      & $270\pm 13$
\end{tabular}
\end{ruledtabular}
\end{table}

\begin{table}
\caption{\label{tab:BrLFQM} The branching ratios for the leptonic
decays of $D$ and $D_s$ in the light-front quark model.}
\begin{ruledtabular}
\begin{tabular}{cccc}
 Decay mode          & Model I  & Model II & Exp.               \\ \hline
 $D_s\to\mu\nu$      & $4.5\times 10^{-3}$ &
 $6.2\times 10^{-3}$ & $(6.2\pm 0.6)\times 10^{-3}$             \\
 $D_s\to\tau\nu$     & $4.4\times 10^{-2}$ &
 $6.0\times 10^{-2}$ & $(6.6\pm 0.6)\times 10^{-2}$             \\
 $D_s\to e\nu$       & $1.1\times 10^{-7}$ &
 $1.5\times 10^{-7}$ & $<1.3\times 10^{-4}$                     \\ \hline
 $D\to\mu\nu$        & $3.6\times 10^{-4}$ &
 $4.4\times 10^{-4}$ & $(4.4\pm 0.7)\times 10^{-4}$             \\
 $D\to\tau\nu$       & $9.6\times 10^{-4}$ &
 $1.2\times 10^{-3}$ & $<2.1\times 10^{-3}$                     \\
 $D\to e\nu$         & $8.5\times 10^{-9}$ &
 $1.0\times 10^{-8}$ & $<2.4\times 10^{-5}$
\end{tabular}
\end{ruledtabular}
\end{table}

Here, we present two different results for decay constants and
branching ratios for semileptonic decays in Table \ref{tab:fLFQM}
and \ref{tab:BrLFQM}. The Model I refers to choosing parameters as
in \cite{Cheng:2003sm}. The decay constants are chosen as $f_D=200$
MeV and $f_{D_s}=230$ MeV. The corresponding parameters are fixed to
$\beta_D=0.448$ GeV, $\beta_{D_s}=0.492$ GeV. Obviously, the decay
constants in this model are smaller than experimental data.  In
Model II, we make the decay constants fit the experiment. In order
to fulfill this, the parameters change to $\beta_D=0.499$ GeV,
$\beta_{D_s}=0.592$ GeV. The experimental data are taken from PDG08
\cite{PDG08}. The ratio of $D\to\tau\nu$ is predicted to be
$1.2\times 10^{-3}$ in Model II, which is  close to the present
experimental upper limit. This process should be observed soon. For
the decays to electron, the ratios are predicted to be of order
$10^{-7}$ or $10^{-8}$ which makes them difficult to observe.

\subsection{Semileptonic decays of $D\to K^{(*)}l\nu$ and $D_s\to
\phi(\eta,\eta')$ in SM}

The semileptonic decays are more complicated in strong dynamics than
leptonic processes because more hadrons are participating. They play
an important role in testing consistency of LFQM approach and new
physics scenarios. After determining parameters $\beta$ from the
leptonic decays, we are able to give predictions for semileptonic
decays. In general, the variations of $\beta$ change the charm meson
wave function and modify the hadron transitions. The chosen
processes  have the same subprocess $c\to sl\nu$ transition as the
leptonic decays $D_s\to l\nu$ since we are interested in the
``$f_{D_s}$ puzzle". Thus, the processes to be considered include
$D\to K^{(*)}l\nu$ and $D_s\to \phi(\eta,\eta')$ decays. They are
classified into two categories: $D\to Pl\nu$ and $D\to Vl\nu$
depending on whether the final meson is pseudoscalar or vector.

For semileptonic decay of D meson to a pseudoscalar, i.e. $D(P)\to
P(P')l\nu$, the differential partial width is given by
\cite{Zweber:2007zz}
 \beq
 \frac{d\Gamma}{dq^2}(D\to Pl\nu)=\frac{G_F^2|V_{cs}|^2~p^3}{24\pi^3}|F_1^{DP}(q^2)|^2,
 \eeq
where $q=P-P'$ is the momentum transfer and $q^2$ is the invariant
mass of the lepton-neutrino pair; $p$ is the final meson momentum in
the D rest frame with
 \beq
 p=|\vec{P'}|=\frac{\sqrt{\left(M_D^2-(M-\sqrt{q^2})^2\right)
  \left(M_D^2-(M+\sqrt{q^2})^2\right)}}{2M_D}.
 \eeq
where $M$ denotes the final meson mass. Neglecting the light lepton
mass, the differential partial width is governed by one form factor
$F_1(q^2)$. The $D\to P$ transition form factors are defined by
 \beq
 \la P(P')|\bar s \gamma_{\mu}c|D(P) \ra=F_{1}(q^2)
  \left[(P+P')_{\mu}-\frac{M_D^2-M^2 }{q^2}q_{\mu}\right]
  +\frac{M_D^2-M^2}{q^2}F_{0}(q^2)\,q_{\mu}.
 \eeq
Then, the total width is
 \beq
 \Gamma(D\to Pl\nu)=\int_0^{(M_D-M)^2}dq^2 \,\frac{d\Gamma}{dq^2}.
 \eeq

For $D(P)\to V(P')l^+\nu$ decays where $V$ represents a vector
meson, the differential partial width is given by
 \beq\label{PVG1}
 \frac{d\Gamma}{dq^2}(D\to Vl\nu)=\frac{G_F^2|V_{cs}|^2}{96\pi^3}
  \frac{p~q^2}{M_D^2}~\sum_{i=+,-,0}~|H_i(q^2)|^2,
 \eeq
where $p$ is the vector meson momentum in the D rest frame;  the
helicity amplitudes $H_i(q^2)$ are given by the combinations of form
factors
 \beq\label{PVG2}
 H_{\pm}(q^2)&=&(M_D+M_V)A_1(q^2)\mp \frac{M_D~p}{M_D+M_V}V(q^2), \non\\
 H_0(q^2)&=&\frac{1}{2M_V\sqrt{q^2}}\Big[(M_D^2-M_V^2-q^2)(M_D+M_V)A_1(q^2)
  -2\frac{p^2 M_D^2}{M_D+M_V}A_2(q^2)\Big].
 \eeq
The $D\to V$ form factors are defined by
 \beq
 \la V(P',\epsilon)|\bar s \gamma_{\mu}c|D(P) \ra &=&
  -\frac{1}{M_D+M_V}\epsilon_{\mu\nu\alpha\beta}
  \epsilon^{*\nu}(P'+P)^{\alpha}q^\beta V(q^2), \non\\
 \la V(P',\epsilon)|\bar s \gamma_{\mu}\gamma_5c|D(P) \ra &=&
 i \Big\{(M_D+M_V)\epsilon^*_\mu A_1(q^2)
  -\frac{\epsilon^*\cdot (P+P')}{M_D+M_V}(P+P')_\mu A_2(q^2)\non\\
  &&-2M_V\frac{\epsilon^*\cdot (P+P')}{q^2}q_\mu
  [A_3(q^2)-A_0(q^2)]\Big\}.
 \eeq
where $\epsilon_\mu$ is the polarization vector of the vector meson
$V$ and it satisfies $\epsilon\cdot P'=0$.

For $P\to P$ and $P\to V$ transition form factors, the detailed
formulas in the covariant light-front approach are given in
\cite{Jaus:1999zv,Cheng:2003sm}. We will not display their explicit
forms here for simplicity. Besides some parameters mentioned in the
previous subsection, other necessary parameters are: $m_s=0.37$ GeV,
$\beta_K=0.3864$ GeV, $\beta_{K^*}=0.2727$ GeV, and
$\beta_\phi=0.3070$ GeV. They are all taken from
\cite{Cheng:2003sm}.

For $\eta$ and $\eta'$ mesons, the case is complicated by their
mixing, i.e., the flavor eigenstates are not the physical states.
Following \cite{Feldmann:1998}, the $\eta-\eta'$ mixing is given by
 \begin{eqnarray}\label{as19}
 \left ( \begin{array}{ccc} \eta \\  \eta' \end{array} \right )=
 \left ( \begin{array}{ccc}
  \rm cos\phi & \rm -sin\phi \\
  \rm sin\phi & \rm cos\phi \end{array} \right )
 \left ( \begin{array}{ccc} \eta_q\\  \eta_s \end{array}\right )
 \end{eqnarray}
where $\phi$ is the mixing angle;
$\eta_q=\frac{u\bar{u}+d\bar{d}}{\sqrt{2}}$ and $\eta_s=s\bar s$.
From the Feynman diagram, we know that only $s\bar{s}$ contributes
to the final meson $\eta(\eta')$ in $D_s \to \eta(\eta')$. We need
to know $\beta^s_\eta$ and $\beta^s_{\eta'}$ in order to calculate
the relevant form factors. In the light-front quark model, the
parameter $\beta$ is extracted from the decay constant. The decay
constants are taken to be $f^s_\eta=-113$ MeV and $f^s_{\eta'}=141$
MeV \cite{Ali:1998}. The  mixing angle is chosen as an averaged
value: $\phi=39.3^\circ$ \cite{Feldmann:1998}. From these above
parameters, we obtain $\beta^s_\eta=0.4059$ GeV and
$\beta^s_{\eta'}=0.4256$ GeV.

In the covariant light-front approach, the formulas of form factors
are derived in the frame $q^+=0$ with $q^2=-q^2_{\perp}\leq 0$, only
values of the form factors in the spacelike momentum region can be
obtained. The advantage of this choice is that the so-called Z-graph
contribution arising from the nonvalence quarks vanishes. In order
to obtain the physical form factors, an analytical extension from
the spacelike region to the timelike region is required. The form
factors in the spacelike region can be parametrized in a
three-parameter form as
 \begin{eqnarray}\label{s14}
 F(q^2)=\frac{F(0)}{
 1-a\left(\frac{q^2}{M_{D}^2}\right)
  +b\left(\frac{q^2}{M_{D}^2}\right)^2}.
 \end{eqnarray}
where $F$ represents the form factors $F_1,~ A_1,~ A_2$ and $V$;
$F(0)$ represents form factor at $q^2=0$. The parameters $F(0)$ and
$a,~b$ are fixed by performing a three-parameter fit to the form
factors in the spacelike region which can be calculated. We then use
these parameters to determine the physical form factors in the
timelike region. The parameters of $a,~b$ and $F(0)$ are fitted from
the form factors at momentum region $-15~{\rm GeV}^2\leq q^2\leq 0$.

\begin{table}
\caption{ The Form Factors of $D$ and $D_s$ in the light-front quark
model.}
\begin{ruledtabular}\label{tab:fLFQM1}
\begin{tabular}{c|ccc||ccc|}
   &              &Model~ I & &        &Model~ II  &         \\\hline
 $F$              &$F(0)$ &a  &b~      &$F(0)$ &a  &b~       \\\hline
 $F_1^{DK}$       &0.79 &1.18 &0.27~~  &0.78 &1.15 &0.24~~   \\\hline
 $A_1^{DK^*}$     &0.65 &0.55 &0.03~~  &0.64 &0.49 &0.02~~   \\\hline
 $A_2^{DK^*}$     &0.57 &1.07 &0.33~~  &0.57 &1.05 &0.27~~   \\\hline
 $V^{DK^*}$       &0.95 &1.35 &0.49~~  &0.90 &1.28 &0.40~~   \\\hline
 $F_1^{D_sK}$     &0.72 &1.27 &0.37~~  &0.70 &1.18 &0.28~~   \\\hline
 $A_1^{D_sK^*}$   &0.56 &0.67 &0.09~~  &0.53 &0.55 &0.04~~   \\\hline
 $A_2^{D_sK^*}$   &0.49 &1.14 &0.50~~  &0.50 &1.08 &0.35~~   \\\hline
 $V^{D_sK^*}$     &0.89 &1.49 &0.76~~  &0.81 &1.34 &0.51~~   \\\hline
 $F_1^{D_s\eta}$  &0.50 &1.17 &0.34~~  &0.48 &1.11 &0.25~~   \\\hline
 $F_1^{D_s\eta'}$ &0.62 &1.14 &0.31~~  &0.60 &1.08 &0.23~~   \\\hline
 $A_1^{D_s\phi}$  &0.65 &0.60 &0.05~~  &0.62 &0.49 &0.02~~   \\\hline
 $A_2^{D_s\phi}$  &0.57 &1.04 &0.37~~  &0.58 &0.99 &0.26~~   \\\hline
 $V^{D_s\phi}$    &1.03 &1.35 &0.57~~  &0.94 &1.22 &0.39~~   \\
\end{tabular}
\end{ruledtabular}
\end{table}

The fitted values of $F(0)$ and $a,~b$ for different form factors
$F_1,~ A_1,~ A_2$ and $V$ are given in Table \ref{tab:fLFQM1}.
Because $A_0$ and $A_3$ don't appear in Eqs. (\ref{PVG1},
\ref{PVG2}), we will not include them in Table \ref{tab:fLFQM1}. Our
results in Model I are consistent with those in \cite{Cheng:2003sm}.
The form factors for $F_1,~ A_1,~ A_2$ in Model I and Model II are
nearly equal. There is only a 5-10\% difference for form factors
$V(q^2)$ in Models I and II.

\begin{table}
\caption{ The branching ratios of semileptonic decays of $D$ and
 $D_s$ in the light-front quark model (in units of \%).}
\begin{ruledtabular}\label{tab:fLFQM2}
\begin{tabular}{cccc}
 decay mode & Model I & Model II & Exp.                     \\\hline
 $D^0\to K^-e^+\nu_e$        &$3.90$ &$3.81$ &$3.58\pm0.06$ \cite{PDG08}    \\\hline
 $D^0\to K^{*-}e^+\nu_e$     &$2.57$ &$2.38$ &$2.38\pm0.16$ \cite{PDG08}    \\\hline
 $D^+\to\bar{K^0}e^+\nu_e$   &$9.96$ &$9.74$ &$8.6 \pm0.5$  \cite{PDG08}    \\\hline
 $D^+\to\bar K^{*0}e^+\nu_e$ &$6.50$ &$6.02$ &$3.66\pm0.21$ \cite{PDG08}    \\\hline\hline
 $D_s^+\to\eta e^+\nu_e$     &$2.42$ &$2.25$ &$2.48\pm0.29\pm0.13$ \cite{Ds}\\\hline
 $D_s^+\to\eta'e^+\nu_e$     &$0.95$ &$0.91$ &$0.91\pm0.33\pm0.05$ \cite{Ds}\\\hline
 $D_s^+\to\phi e^+\nu_e$     &$2.95$ &$2.58$ &$2.29\pm0.37\pm0.11$ \cite{Ds}
\end{tabular}
\end{ruledtabular}
\end{table}

Now, we give predictions for branching ratios of semileptonic decays
of $D$ and $D_s$. The results are displayed in Table
\ref{tab:fLFQM2}. The experimental data about $D_s$ decays are taken
from a most recent measurement from CLEO collaborations \cite{Ds}.
About the numerical results, some comments are given below:

(1) The predictions in Model II are in general smaller than those in
Model I. For most processes, the results in Model II are closer to
the experimental data. In other words, the semileptonic decays
prefer larger decay constants of $D$ and $D_s$ which is indicated by
leptonic processes.

(2) For $D_s^+\to\eta(\eta') e^+\nu_e$, $D_s^+\to\phi e^+\nu_e$ and
$D^0\to K^{*-}e^+\nu_e$ decays, the theory is in good agreement with
the experiment.

(3) It is difficult to understand the decay $D^+\to\bar
K^{*0}e^+\nu_e$ where the theory prediction is larger than the
experiment. Under the isospin symmetry, $\frac{Br(D^+\to\bar
K^{*0}e^+\nu_e)}{Br(D^0\to
K^{*-}e^+\nu_e)}=\frac{\tau_{D^+}}{\tau_{D^0}}=2.54$. But the
experiment result is $\frac{Br(D^+\to\bar K^{*0}e^+\nu_e)}{Br(D^0\to
K^{*-}e^+\nu_e)}=\frac{\tau_{D^+}}{\tau_{D^0}}=1.54$. The
discrepancy between the theory and experiment may be related to the
old puzzle about life time difference between $D^+$ and $D^0$.

(4) Our results favor a large $\eta-\eta'$ mixing angle
$\phi=39^\circ$. This may be due to our neglecting the glue
component in $\eta'$.

\section{Leptonic and semileptonic decays of $D$ and $D_s$ mesons
 in unparticle physics}

\subsection{Leptonic decays in unparticle physics}

As discussed in the introduction, new physics effects must interfere
constructively with the SM contribution and enhance the  rate of
leptonic decay. Unparticle physics can provide such interference due
to the nontrivial phase effects. The scale dimension $d_\Un$ of the
unparticle is in general fractional rather than an integral number.
The fractional dimension induce a phase factor $e^{-id_\Un\pi}$ in
the propagator of the unparticle field.

The scale invariant unparticle fields emerge below an energy scale
$\Lambda_\Un$ which is at the order of TeV. The unparticle has some
peculiar characteristics that make it different from the ordinary
particle. The interactions between the unparticle and the SM
particles are described in the framework of low energy effective
theory. For our purpose, the coupling of a scalar unparticle to two
SM fermions (quarks or leptons) is given by an effective interaction
as
 \beq
 {\cal L}_{\rm eff}^{\Un}=\frac{C_{f'f}}{\Lambda_\Un^{d_\Un}}
  \bar f'\gamma_{\mu}(1-\gamma_5)f\partial^\mu O_\Un+h.c. .
 \eeq
where $O_\Un$ denotes the scalar unparticle fields. The $C_{f'f}$
are dimensionless coefficients and they depend on different flavors
in general. There are two reasons that we don't consider the vector
unparticle. (1) The transverse vector unparticle does not contribute
to the leptonic decay of a pseudoscalar meson \cite{Huang:2007ax}.
(2) Even if there is a nontransverse vector contribution, another
constraint will suppress it significantly. For the vector
unparticle, it is pointed out that conformal symmetry puts a lower
bound on its scale dimension $d_\Un\geq 3$ \cite{Grinstein:2008qk}.
If we take this constraint seriously, the vector unparticle effects
in most processes will be very small and are negligible.

In this study, we are only interested in the effects of the virtual
unparticle, thus it only appears as a propagator with momentum $P$
and scale dimension $d_\Un$.  The propagator for the scalar
unparticle field in the timelike momentum region with $P^2\geq 0$ is
given by \cite{Georgi:2007si,Cheung:2007zza}
 \beq
 \int d^4 x e^{iP\cdot x}\la 0 |TO_\Un(x)O_\Un(0)|0\ra &=&
   i\frac{A_{d_\Un}}{2~{\rm sin}(d_\Un\pi)}\frac{1}{(P^2+i\epsilon)^{2-d_\Un}}
   e^{-id_\Un\pi},
 \eeq
where
 \beq
 A_{d_\Un}=\frac{16\pi^{5/2}}{(2\pi)^{2d_\Un}}\frac{\Gamma(d_\Un+1/2)}
  {\Gamma(d_\Un-1)\Gamma(2d_\Un)}.
 \eeq
The function ${\rm sin}(d_\Un\pi)$ in the denominator implies that
the scale dimension $d_\Un$ cannot be integral for $d_\Un>1$ in
order to avoid singularity (for $d_\Un=1$ the singularity is
canceled by $\Gamma(d_\Un-1)$ term in $A_{d_\Un}$). The phase factor
$e^{-id_\Un\pi}$ provides a CP conserving phase which produces
peculiar interference effects in high energy scattering processes,
CP violation in B and D decays, etc.. There may be scale symmetry
violation after the spontaneous breaking due to the coupling of the
unparticle to the Higgs \cite{Fox:2007sy}. The estimate of the
violation is difficult, so we neglect it in this study.

\begin{figure}[!htb]
\begin{center}
\begin{tabular}{cc}
\includegraphics[width=7cm]{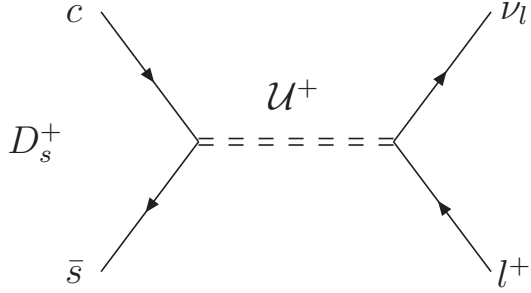}
\end{tabular}
\end{center}
\caption{ The lowest order diagram for the decay of $D_s^+\to
l^+\nu_l$ in unparticle physics. The double dashed lines represent
the unparticle.} \label{fig3}
\end{figure}

The Feynman diagram of unparticle contribution is obtained by
replacing the W boson to unparticle field. Fig. \ref{fig3} depicts
the lowest order diagram for the decay of $D_s^+\to l^+\nu_l$ in
unparticle physics. The amplitude for quark level transition of
$c\to ql\nu$ is
 \beq \label{eq:AUn}
 A^\Un=-i\frac{A_{d_\Un}}{2~{\rm sin}(d_\Un\pi)}\frac{C_q}
  {\Lambda_\Un^{2d_\Un}}\frac{P^\mu P^\nu}{(P^2)^{2-d_\Un}}e^{-id_\Un\pi}~
  \bar q\gamma_\mu(1-\gamma_5)c~\bar\nu\gamma_\nu(1-\gamma_5)l,
 \eeq
where $P^2=M_D^2$ in $D\to l\nu$ decays and $C_q\equiv
C_{cq}C_{l\nu}$. In the SM, the W boson can be integrated out and
the interaction of four fermions becomes a local interaction at low
energy. Because unparticle is different from a heavy particle with a
fixed mass, the unparticle propagation is a non-local interaction.
But $P^2$ is a constant, we can still consider the $c\to q l\nu$
transition as an effective interaction. Considering Eq.
(\ref{eq:fP}), the $P^\mu P^\nu$ term in Eq. (\ref{eq:AUn}) can be
replaced by $P^2g^{\mu\nu}$ (note that they are not equal) in the
final result. The $c\to q l\nu$ transition is rewritten by
 \beq \label{eq:HUn}
 {\cal H}_{\rm eff}^{\Un}&=&\frac{A_{d_\Un}}{2~{\rm sin}(d_\Un\pi)}
  \frac{C_q}{M_D^2}\left(\frac{M_D^2}{\Lambda_\Un^2}\right)^{d_\Un}
  e^{-id_\Un\pi}~(\bar q c)_{V-A}(\bar\nu l)_{V-A}
  =re^{-i\phi_\Un}~{\cal H}_{\rm eff}^{SM},
 \eeq
where $r$ and $\phi_\Un$ are
 \beq
 &&r=\frac{A_{d_\Un}}{2~{\rm sin}(d_\Un\pi)}\frac{C_q}{M_D^2} \left(
 \frac{M_D^2}{\Lambda_\Un^2}\right)^{d_\Un}\frac{\sqrt{2}}{G_F V_{cq}}\,,\non\\
 &&\phi_\Un=d_\Un\pi.
 \eeq
Eq. (\ref{eq:HUn}) shows a clear physical meaning: the unparticle
effects are equivalent to multiplying a constant factor with a
CP-conserving phase to the SM contribution.

Combining the SM and unparticle contributions, we obtain the decay
rate of $D\to l\nu$ decays as
 \beq
 \Gamma(D\to l\nu)=\Gamma^{\rm SM}(D\to l\nu)
  \left|1+re^{-i\phi_\Un}\right|^2.
 \eeq
Our result is principally consistent with the formulation for B
decays in \cite{Zweber:2007zz,Huang:2007ax}.

In \cite{Zweber:2007zz,Huang:2007ax}, the authors point out a novel
CP asymmetry in the leptonic decays caused by the CP-even phase of
unparticle. The phase $\phi_\Un$ mimics the strong interaction phase
caused by final state interactions.  This phenomenon distinguishes
unparticle model from other new physics scenarios. If there is CP
asymmetry in $D\to l\nu$ decays observed in experiment, it would be
a clear signal of unparticle physics. Unlike the $B^+$ decay where
the CKM parameter $V_{ub}$ contains a CP violating weak phase, the
$V_{cd}$ and $V_{cs}$ in $D$ decays have no weak phase in SM. In
order to produce CP asymmetry which requires both CP-even and CP-odd
phase differences, we make an assumption that the coupling
coefficient $C_{cq}C_{l\nu}$ is complex and contains a CP-odd phase
$\phi_w$ even if its origin is unknown, and then $C_{cq}C_{l\nu}\to
C_{cq}C_{l\nu}e^{-i\phi_w}$. After this assumption, the direct CP
asymmetry in $D_s\to l\nu$ decay is
 \beq
 A_{CP}(D_s\to l\nu)&\equiv&
  \frac{\Gamma(D_s^-\to l^-\bar\nu)-\Gamma(D_s^+\to l^+\nu)}
  {\Gamma(D_s^-\to l^-\bar\nu)+\Gamma(D_s^+\to l^+\nu)}\non\\
  &=&\frac{2r{\rm sin}\phi_\Un {\rm sin}\phi_w}
   {1+r^2+2r{\rm cos}\phi_\Un {\rm cos}\phi_w}\,.
 \eeq
The CP violation is caused by the interference between SM and
unparticle contributions.

\begin{figure}
\includegraphics[width=10cm]{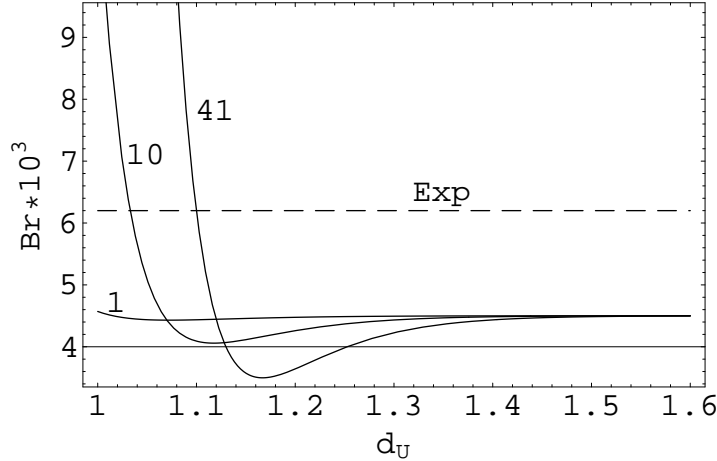}
\caption{ The branching ratio of $D_{s}\to \mu \nu_{\mu}$ vs the
 scale dimension $d_\Un$. The $C_s$ are taken to be three values:
 1, 10, 41.}\label{fDsBrUn}
\end{figure}

\begin{figure}
\includegraphics[width=10cm]{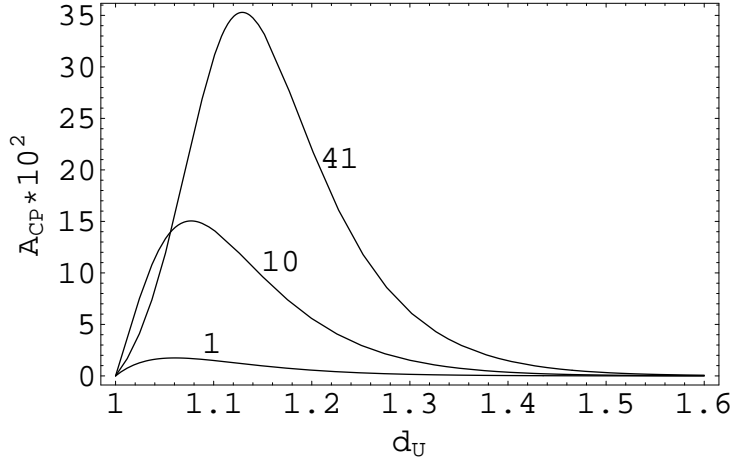}
\label{figure} \caption{The direct CP asymmetry of $D_{s}\to \mu
 \nu_{\mu}$ vs the scale dimension $d_\Un$. The $C_s$ are taken to be:
 1, 10, 41.}\label{fig5}
\end{figure}

\begin{figure}
\includegraphics[width=10cm]{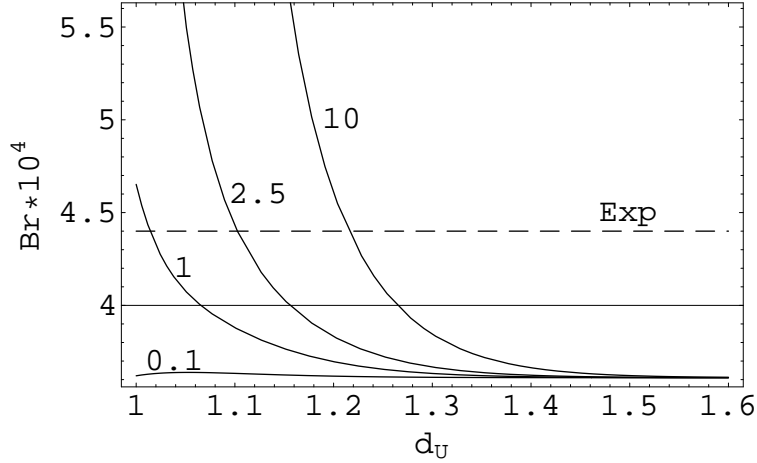}
\caption{ The branching ratio of $D\to \mu \nu_{\mu}$ vs the
 scale dimension $d_\Un$. The $C_d$ are taken to be four values:
 0.1, 1, 2.5, 10.}\label{fig6}
\end{figure}

\begin{figure}
\includegraphics[width=10cm]{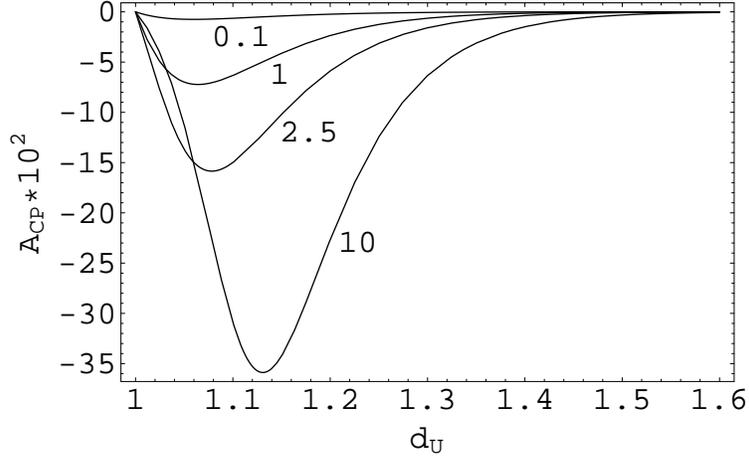}
\label{figure} \caption{The direct CP asymmetry of $D\to \mu
 \nu_{\mu}$ vs the scale dimension $d_\Un$. The $C_d$ are taken to be:
 0.1, 1, 2.5, 10.}\label{fig7}
\end{figure}

Now, we discuss the numerical results. We fix the scale parameter
$\Lambda_\Un=1$ TeV. The coupling coefficients have been defined to
be $C_{s}=C_{cs}C_{l\nu}$ and $C_{d}=C_{cd}C_{l\nu}$. The SM
parameters are taken from Model I: $f_D=200$ MeV, $f_{D_s}=230$ MeV.
We don't use Model II since new physics is not necessary in it. At
first, we give the results for $D_s\to \mu \nu_{\mu}$ where new
physics effect is expected to be most important. Fig.\ref{fDsBrUn}
plots the dependence of branching ratio on the scale dimension
$d_\Un$ at different values of $C_s=1,~10,~41$. We gives the results
in the range $1<d_\Un<1.6$. In order to explain the experiment, the
$C_s$ needs to be larger than 10. For the CP asymmetry, we consider
a maximal case, i.e. the CP-odd phase $\phi_w=\pi/2$. Fig.
\ref{fig5} plots the dependence of direct CP asymmetry on the scale
dimension $d_\Un$ also at values of $C_s=1,~10,~41$. It is seen that
the maximal $A_{CP}$ can reach 35\%. Considering the experimental
constraint for branching ratio, $A_{CP}$ reaches 10\% for $C_s=10$
and 30\% for $C_s=44$. Thus, the order 10\% CP asymmetry is possible
in unparticle physics. Our predictions for direct CP violation seems
large. This is because we adopt a maximal case for the CP-odd phase.
This phase is unknown and other choices will decrease the
predictions. A recent measurement from CLEO Collaboration shows no
CP violation, $A_{CP}=(4.8\pm 6.1)\%$ \cite{Alexander:2009ux}. But
it does not exclude the unparticle scenario because the experimental
errors are large and we consider the maximal CP violation in theory.
In fact, even a $1\%$ CP violation is a support for unparticle
theory. Similar results for $D\to \mu \nu_{\mu}$ decay are given in
Figs. \ref{fig6} and \ref{fig7}.

The experiments provide stringent constraints on the unparticle
parameters. Without loss of generality, scale dimension is chosen
to be $d_\Un=1.1$. Then the observed branching ratios of
$D_s(D)\to l\nu$ can be used to constrain the  coupling
coefficients $C_d$ and $C_s$. Table \ref{tab:Csd} lists the
constraints on the $C_{s}$ and $C_{d}$ from $D_s(D)\to l\nu$
decays.

\begin{table}
\caption{The constraints on the $C_{s}$ and $C_{d}$ from
 $D_s(D)\to l\nu$ decays with $d_\Un=1.1$. } \label{tab:Csd}
\begin{ruledtabular}
 \begin{tabular}{cccc}
         & $D_s\to\mu\nu$ & $D_s\to\tau\nu$  & $D_s\to e\nu$ \\\hline
 $C_{s}$ & $41$           & $46$             & -             \\\hline\hline
         & $D\to\mu\nu$   & $D\to\tau\nu$    & $D\to e\nu$   \\\hline
 $C_{d}$ & $2.5$          & -                & -             %\\\hline\hline
\end{tabular}
\end{ruledtabular}
\end{table}

\subsection{Semileptonic decays in unparticle physics}

For the semileptonic decays, the quark level transition of $c\to
ql\nu$ is nearly the same as that in the leptonic decay except the
momentum of the unparticle is not equal to that of D meson. The
unparticle momentum is equal to the lepton pair momentum $q$, and
the amplitude of the subprocess is
 \beq
 A^\Un=-i\frac{A_{d_\Un}}{2~{\rm sin}(d_\Un\pi)}\frac{C_q}
  {\Lambda_\Un^{2d_\Un}}\frac{q^\mu q^\nu}{(q^2)^{2-d_\Un}}e^{-id_\Un\pi}~
  \bar q\gamma_\mu(1-\gamma_5)c~\bar\nu\gamma_\nu(1-\gamma_5)l.
 \eeq
By using the equation of motion,
$\bar\nu\gamma_\nu(1-\gamma_5)l~q^\nu=0$ in the zero lepton mass
limit. Thus, a conclusion is obtained: the scalar unparticle
contribution is helicity suppressed and vanishes for
semileptonically decaying to the light lepton ($e,~\mu$). In the
leptonic decay case, because the leading SM contribution suffers the
helicity suppression, the unparticle effects, although suppressed,
play an important role. While for the semileptonic decay, the
leading SM contribution is not suppressed. So the unpaticle effects
are negligible due to helicity suppression and the weak coupling
with SM particles. The vector unparticle may contribute to the
semileptonic decays, but it does not enhance the ratio of the
leptonic decay and cannot solve the $f_{D_s}$ puzzle.

\section{Discussions and conclusions}

We have studied the leptonic and semileptonic decays of $D$ and
$D_s$ within SM utilizing a light-front quark model. We find that it
is not difficult to solve  the $f_{D_s}$ puzzle by adjusting
parameter reasonably. The predictions for semileptonic decays are
consistent with experiment. Although the numerical results depend on
the model and the theory uncertainties are not under control, the
conclusion may be general and model independent. There is a
sufficient space due to strong interaction uncertainties which can
explain the discrepancy between theory and experiment. This
conclusion is different from claims from lattice QCD and QCD sum
rules.

We also study the leptonic decays in unparticle physics. The
unparticle induces  constructive interference effects which can
enhance the theory to be consistent with the experiment. Production
of CP asymmetry at percent level is possible. This would be a clear
signal for unparticle physics. We hope the future experiment can
test its validity.

The discrepancy between the theory and experiment becomes smaller if
we use the most recent measurement from CLEO collaboration with
$f_{D_s}=259.5\pm 6.6\pm 3.1$ MeV \cite{Alexander:2009ux}. The
solution of the $f_{D_s}$ puzzle requires more accurate measurements
on the leptonic and semileptonic decays of charm mesons. The future
BESIII will provide precise determination of $D_s$ decay constant
with errors to 1\% \cite{Li:2008wv}.

In conclusion, there is space in SM to interpret the $f_{D_s}$
puzzle without contradictions with the other experiments. The
observation of CP violation at percent level may be an ideal test of
the unparticle scenario.

\section*{Acknowledgments}

This work was supported in part by the National Natural Science
Foundation of China (NNSFC) under Contract No. 10705015.


\begin{thebibliography}{99}

\bibitem{Bigi:2009df} For some recent reviews, see:
  I.~I.~Bigi, M.~Blanke, A.~J.~Buras and S.~Recksiegel,
  %``CP Violation in D0 - anti-D0 Oscillations: General Considerations and
  %Applications to the Littlest Higgs Model with T-Parity,''
  arXiv:0904.1545 [hep-ph];
%
  X.~Q.~Li, X.~Liu and Z.~T.~Wei,
  %``Charm Physics: A Field Full with Challenges and Opportunities,''
  Front.\ Phys.\ China {\bf 4}, 49 (2009)
  [arXiv:0808.2587 [hep-ph]].
  %%CITATION = 00515,4,49;%%

%\cite{Rosner:2008yu}
\bibitem{Rosner:2008yu}
  J.~L.~Rosner and S.~Stone,
  %``Decay Constants of Charged Pseudoscalar Mesons,''
  arXiv:0802.1043 [hep-ex].
  %%CITATION = ARXIV:0802.1043;%%

%\cite{Follana:2007uv}
\bibitem{Follana:2007uv}
  E.~Follana, C.~T.~H.~Davies, G.~P.~Lepage and J.~Shigemitsu  [HPQCD
                  Collaboration and UKQCD Collaboration],
  %``High Precision determination of the pi, K, D and D_s decay constants   from
  %lattice QCD,''
  Phys.\ Rev.\ Lett.\  {\bf 100}, 062002 (2008)
  [arXiv:0706.1726 [hep-lat]].
  %%CITATION = PRLTA,100,062002;%%

%\cite{Khodjamirian:2008xt}
\bibitem{Khodjamirian:2008xt}
  A.~Khodjamirian,
  %``Upper bounds on $f_D$ and $f_{D_s}$ from two-point correlation function in
  %QCD,''
  Phys.\ Rev.\  D {\bf 79}, 031503 (2009)
  [arXiv:0812.3747 [hep-ph]].
  %%CITATION = PHRVA,D79,031503;%%

%\cite{Jaus:1999zv}
\bibitem{Jaus:1999zv}
  W.~Jaus,
  %``Covariant analysis of the light-front quark model,''
  Phys.\ Rev.\  D {\bf 60}, 054026 (1999);
  %%CITATION = PHRVA,D60,054026;%%

%\cite{Cheng:2003sm}
\bibitem{Cheng:2003sm}
  H.~Y.~Cheng, C.~K.~Chua and C.~W.~Hwang,
  %``Covariant light-front approach for s-wave and p-wave mesons: Its
  %application to decay constants and form factors,''
  Phys.\ Rev.\  D {\bf 69}, 074025 (2004)
  [arXiv:hep-ph/0310359].
  %%CITATION = PHRVA,D69,074025;%%

%\cite{Akeroyd:2007eh}
\bibitem{Akeroyd:2007eh}
  A.~G.~Akeroyd and C.~H.~Chen,
  %``Effect of H^\pm on B^\pm\to \tau^\pm\nu_\tau and D^\pm_s\to
  %\mu^\pm\nu_\mu,\tau^\pm\nu_\tau,''
  Phys.\ Rev.\  D {\bf 75}, 075004 (2007)
  [arXiv:hep-ph/0701078].
  %%CITATION = PHRVA,D75,075004;%%

%\cite{Akeroyd:2009tn}
\bibitem{Akeroyd:2009tn}
  A.~G.~Akeroyd and F.~Mahmoudi,
  %``Constraints on charged Higgs bosons from D(s)+- -> mu+- nu and D(s)+- ->
  %tau+- nu,''
  JHEP {\bf 0904}, 121 (2009)
  [arXiv:0902.2393 [hep-ph]].
  %%CITATION = JHEPA,0904,121;%%

%\cite{Dobrescu:2008er}
\bibitem{Dobrescu:2008er}
  B.~A.~Dobrescu and A.~S.~Kronfeld,
  %``Accumulating evidence for nonstandard leptonic decays of $D_s$ mesons,''
  Phys.\ Rev.\ Lett.\  {\bf 100}, 241802 (2008)
  [arXiv:0803.0512 [hep-ph]].
  %%CITATION = PRLTA,100,241802;%%

%\cite{Benbrik:2008ik}
\bibitem{Benbrik:2008ik}
  R.~Benbrik and C.~H.~Chen,
  %``Leptoquark on $P\to \ell^{+} \nu$, FCNC and LFV,''
  Phys.\ Lett.\  B {\bf 672}, 172 (2009)
  [arXiv:0807.2373 [hep-ph]].
  %%CITATION = PHLTA,B672,172;%%



%\cite{Georgi:2007ek}
\bibitem{Georgi:2007ek}
  H.~Georgi,
  %``Unparticle Physics,''
  Phys.\ Rev.\ Lett.\  {\bf 98}, 221601 (2007)
  [arXiv:hep-ph/0703260].
  %%CITATION = PRLTA,98,221601;%%

%\cite{Georgi:2007si}
\bibitem{Georgi:2007si}
  H.~Georgi,
  %``Another Odd Thing About Unparticle Physics,''
  Phys.\ Lett.\  B {\bf 650}, 275 (2007)
  [arXiv:0704.2457 [hep-ph]].
  %%CITATION = PHLTA,B650,275;%%

%\cite{Cheung:2007zza}
\bibitem{Cheung:2007zza}
  K.~Cheung, W.~Y.~Keung and T.~C.~Yuan,
  %``Collider signals of unparticle physics,''
  Phys.\ Rev.\ Lett.\  {\bf 99}, 051803 (2007)
  [arXiv:0704.2588 [hep-ph]].
  %%CITATION = PRLTA,99,051803;%%

\bibitem{Unparticle}
 %\cite{Li:2007by}
 %\bibitem{Li:2007by}
  X.~Q.~Li and Z.~T.~Wei,
  %``Unparticle physics effects on D0 - anti-D0 mixing,''
  Phys.\ Lett.\  B {\bf 651}, 380 (2007)
  [arXiv:0705.1821 [hep-ph]];
  %%CITATION = PHLTA,B651,380;%%
 %\cite{Li:2007kj}
 %\bibitem{Li:2007kj}
  X.~Q.~Li, Y.~Liu and Z.~T.~Wei,
  %``Neutrino decay as a possible interpretation to the MiniBooNE observation
  %with unparticle scenario,''
  Eur.\ Phys.\ J.\  C {\bf 56}, 97 (2008)
  [arXiv:0707.2285 [hep-ph]];
  %%CITATION = EPHJA,C56,97;%%
%
 %\cite{Liu:2007twb}
 %\bibitem{Liu:2007twb}
  X.~Liu, H.~W.~Ke, Q.~P.~Qiao, Z.~T.~Wei and X.~Q.~Li,
  %``A Possibility of Search for New Physics at LHCb,''
  Phys.\ Rev.\  D {\bf 77}, 035014 (2008)
  [arXiv:0710.2600 [hep-ph]];
  %%CITATION = PHRVA,D77,035014;%%
 %\cite{Chen:2007cz}
%
  %\bibitem{Chen:2007cz}
  S.~L.~Chen, X.~G.~He, X.~Q.~Li, H.~C.~Tsai and Z.~T.~Wei,
  %``Constraints on Unparticle Interactions from Particle and Antiparticle
  %Oscillations,''
  Eur.\ Phys.\ J.\  C {\bf 59}, 899 (2009)
  [arXiv:0710.3663 [hep-ph]];
  %%CITATION = EPHJA,C59,899;%%
%
 %\cite{Wei:2008wi}
 %\bibitem{Wei:2008wi}
  Z.~T.~Wei, Y.~Xu and X.~Q.~Li,
  %``Probing unparticle theory via lepton flavor violating process $J/\psi\to
  %ll'$ at BESIII,''
  arXiv:0806.2944 [hep-ph];
  %%CITATION = ARXIV:0806.2944;%%
%
 %\cite{Wei:2008zzc}
 %\bibitem{Wei:2008zzc}
  Z.~T.~Wei,
  %``Unparticle physics and neutral meson mixing,''
  Int.\ J.\ Mod.\ Phys.\  A {\bf 23}, 3339 (2008).
  %%CITATION = IMPAE,A23,3339;%%


%\cite{Huang:2007ax}
\bibitem{Huang:2007ax}
  C.~S.~Huang and X.~H.~Wu,
  %``Direct CP violation of $B \to l \nu$ in unparticle physics,''
  Phys.\ Rev.\  D {\bf 77}, 075014 (2008)
  [arXiv:0707.1268 [hep-ph]].
  %%CITATION = PHRVA,D77,075014;%%

%\cite{Zwicky:2007vv}
\bibitem{Zwicky:2007vv}
  R.~Zwicky,
  %``Unparticles at heavy flavour scales: CP violating phenomena,''
  Phys.\ Rev.\  D {\bf 77}, 036004 (2008)
  [arXiv:0707.0677 [hep-ph]].
  %%CITATION = PHRVA,D77,036004;%%


\bibitem{PDG08}
  C.~Amsler {\it et al.}  [Particle Data Group],
  %``Review of particle physics,''
  Phys.\ Lett.\  B {\bf 667}, 1 (2008).


%\cite{Hwang:2006cua}
\bibitem{Hwang:2006cua}
  C.~W.~Hwang and Z.~T.~Wei,
  %``Covariant light-front approach for heavy quarkonium: decay constants, P \to
  %\gamma \gamma and V \to P \gamma,''
  J.\ Phys.\ G {\bf 34}, 687 (2007)
  [arXiv:hep-ph/0609036];
  %%CITATION = JPHGB,G34,687;%%
%\cite{Lu:2007sg}
%\bibitem{Lu:2007sg}
  C.~D.~Lu, W.~Wang and Z.~T.~Wei,
  %``Heavy-to-light form factors on the light cone,''
  Phys.\ Rev.\  D {\bf 76}, 014013 (2007)
  [arXiv:hep-ph/0701265].
  %%CITATION = PHRVA,D76,014013;%%
%\cite{Ke:2007tg}
%\bibitem{Ke:2007tg}
  H.~W.~Ke, X.~Q.~Li and Z.~T.~Wei,
  %``Diquarks and \Lambda_{b}\to\Lambda_c weak decays,''
  Phys.\ Rev.\  D {\bf 77}, 014020 (2008)
  [arXiv:0710.1927 [hep-ph]];
  %%CITATION = PHRVA,D77,014020;%%

%\cite{Zweber:2007zz}
\bibitem{Zweber:2007zz}
  P.~Zweber,
  %``Charm leptonic and semileptonic decays,''
  Nucl.\ Phys.\ Proc.\ Suppl.\  {\bf 170}, 107 (2007).
  %%CITATION = NUPHZ,170,107;%%



\bibitem{Feldmann:1998}
 Th. Feldmann , P. Kroll and B. Stech,
  %``Experimental tests of factorization in charmless nonleptonic two-body B decays,''
  Phys.\ Rev.\  D {\bf 58}, 114006 (1998).
\bibitem{Ali:1998}
 A. Ali, G. Kramer and C.~D.~L\"{u},
  %``Experimental tests of factorization in charmless nonleptonic two-body B decays,''
  Phys.\ Rev.\  D {\bf 58}, 094009 (1998).

\bibitem{Ds}
 J. Yelton {\it et al.},   [CLEO Collaboration],
  %``Absolute Branching Fraction Measurements for Exclusive $D_s$ Semileptonic
  %Decays,''
  arXiv:0903.0601 [hep-ex].
  %%CITATION = ARXIV:0903.0601;%%


%\cite{Grinstein:2008qk}
\bibitem{Grinstein:2008qk}
 B.~Grinstein, K.~A.~Intriligator and I.~Z.~Rothstein,
 %``Comments on Unparticles,''
 Phys.\ Lett.\  B {\bf 662}, 367 (2008)
 [arXiv:0801.1140 [hep-ph]];
 %\cite{Nakayama:2007qu}
 %\bibitem{Nakayama:2007qu}
  Y.~Nakayama,
  %``SUSY Unparticle and Conformal Sequestering,''
  Phys.\ Rev.\  D {\bf 76}, 105009 (2007)
  [arXiv:0707.2451 [hep-ph]].
  %%CITATION = PHRVA,D76,105009;%%


%\cite{Fox:2007sy}
\bibitem{Fox:2007sy}
  P.~J.~Fox, A.~Rajaraman and Y.~Shirman,
  %``Bounds on Unparticles from the Higgs Sector,''
  Phys.\ Rev.\  D {\bf 76}, 075004 (2007)
  [arXiv:0705.3092 [hep-ph]].
  %%CITATION = PHRVA,D76,075004;%%


%\cite{Alexander:2009ux}
\bibitem{Alexander:2009ux}
  J.~P.~Alexander {\it et al.}  [CLEO Collaboration],
  %``Measurement of $B{D_s^+ \to \ell^+ \nu}$ and the Decay Constant $fD_s^+$
  %From 600 $/pb^{-1}$ of $e^\pm$ Annihilation Data Near 4170 MeV,''
  Phys.\ Rev.\  D {\bf 79}, 052001 (2009)
  [arXiv:0901.1216 [hep-ex]].
  %%CITATION = PHRVA,D79,052001;%%




%\cite{Li:2008wv}
\bibitem{Li:2008wv}
 J. Zou, H. Li and X. Zhang,
  %``A possible signature of new physics at BES-III,''
  arXiv:0804.1822 [hep-ex].
  %%CITATION = ARXIV:0804.1822;%%


\end{thebibliography}
\end{document}